\documentclass[runningheads]{llncs}
\usepackage[T1]{fontenc}
\usepackage{irules}
\usepackage{amsmath}
\usepackage{amsfonts}
\usepackage{microtype}
\usepackage{color}
\usepackage{listings}
\usepackage{hyperref}
\usepackage[colorinlistoftodos,prependcaption,textsize=tiny]{todonotes}

\tikzset{
  > = stealth,
  shorten > = 2pt,
  shorten < = 2pt
}

\usetikzlibrary{
  arrows
}

\definecolor{strings}{rgb}{0,0.5,0}
\definecolor{emphs}{rgb}{0.64,0.08,0.08}
\definecolor{comments}{rgb}{0.17,0.57,0.68}
\colorlet{keywords}{blue!50!cyan}

\lstdefinestyle{tinysol}{
  language=C,
  captionpos=b,
  numbers=left,
  numberstyle=\tiny,
  frame=lines,
  showspaces=false,
  showtabs=false,
  breaklines=true,
  showstringspaces=false,
  breakatwhitespace=true,
  emph={contract,field,func,call,if,then,else,while,do},
  emphstyle={\rmfamily\bfseries\color{emphs}},
  commentstyle=\color{comments},
  morekeywords={contract, interface, account, this, var, field, func, method, send, value, balance, sender, then, fallback, call, dcall, args, id},
  keywordstyle={\bfseries\color{keywords}},
  stringstyle=\color{strings},
  basicstyle=\ttfamily\small,
  escapechar=@
}

\title{Information Flow Control in Off-Chain Components}
\author{%
     Stian Lybech\inst{1}\orcidID{0000-0001-8219-2285}
\and Eun-Young Kang\inst{1}\orcidID{0000-0002-4589-2378} 
\and Riccardo Tonello\inst{2}\orcidID{0000-0001-8098-3235}
\and Anders Dalskov\inst{3}\orcidID{0000-0002-3881-4742}
}

\authorrunning{S. Lybech et al.}

\institute{%
      University of Southern Denmark, Odense, Denmark \email{\{stly,eyk\}@mmmi.sdu.dk} 
\and Danish Technological Institute, Aarhus, Denmark \email{rict@teknologisk.dk}
\and Partisia Applications ApS, Aarhus, Denmark \email{anderspkd@partisia.com}
}

\DeclareTextFontCommand{\syntaxfont}{\normalfont\ttfamily}
\newcommand{\code}[1]{\syntaxfont{#1}}

\newcommand{\ifrelax}[3]{%
\expandafter\ifx#1\relax%
  #2%
\else 
  #3%
\fi
}

\DeclareMathOperator{\op}{op}   
\DeclareMathOperator{\dom}{dom} 

\newcommand{\SKIP}{\code{skip}}

\newcommand{\DECVAR}[3]{ \code{var }#1\code{ := }#2\code{ in }#3 }
\newcommand{\ASSIGN}[2]{ #1\code{ := }#2 }
\newcommand{\IFTHEN}[3]{ \code{if }#1\code{ then }#2\code{ else }#3 }

\newcommand{\FORK}[2]{ \code{fork } #1\code{;} #2 }

\NewDocumentCommand\CALL{ O{\relax} m m m O{\relax} O{\relax} }{%
\ifrelax{#1}{}{\code{#1 }}%
\ifrelax{#6}{}{#6\code{!}}%
#2\code{.}#3\code{(}#4\code{)}%
\ifrelax{#5}{}{\code{:}#5}%
}

\newcommand{\RPCCALL}[5]{ \CALL[call]{#1}{#2}{#3}[#4][#5] }
\newcommand{\LOCCALL}[3]{ \CALL[call]{\code{this}}{#1}{#2}[#3] }

\newcommand{\ENDSCOPE}[1]{ \code{end(}#1\code{)} }
\newcommand{\ENDCALL}[1]{ \code{ret(}#1\code{)} }
\newcommand{\PUB}{\code{pub}}
\newcommand{\CALLBACK}[1]{ \code{cb(}#1\code{)} }
\newcommand{\TRANS}{ \code{trans} }

\newcommand{\TRANSACT}[5]{ #1\code{->}\CALL{#2}{#3}{#4}[#5] }

\newcommand{\LOC}[4]{#1[#2]^{#3}_{#4}}

\newcommand{\ENV}[1]{\ensuremath{env_{#1}}}
\newcommand{\SETENV}[1]{\ensuremath{\mathcal{E}_{\kern-2pt#1}}}
\newcommand{\SETNAMES}[1]{\ensuremath{\mathcal{N}_{\kern-2pt#1}}}

\def\BOOLEANS{\ensuremath{\mathbb{B}}}
\def\EMPTY{\ensuremath{\emptyset}}

\def\EVENTS{\ensuremath{\mathcal{R}}}
\def\VALUES{\ensuremath{\mathcal{V}}}
\def\EXPRESSIONS{\ensuremath{\mathcal{E}}}
\def\STATEMENTS{\ensuremath{\mathcal{S}}}
\def\STACKS{\ensuremath{\mathcal{Q}}}
\def\TRANSACTIONS{\ensuremath{\mathcal{T}}}
\def\THREADPOOLS{\ensuremath{\mathcal{P}}}
\def\LOCATIONS{\ensuremath{\mathcal{L}}}

\def\ANAMES{\SETNAMES{A}} 
\def\VNAMES{\SETNAMES{V}} 
\def\FNAMES{\SETNAMES{F}} 
\def\PNAMES{\SETNAMES{P}} 
\def\DNAMES{\SETNAMES{D}} 

\def\SETENVM{\SETENV{M}} 
\def\SETENVC{\SETENV{C}} 
\def\SETENVV{\SETENV{V}} 

\def\ENVC{\ENV{C}} 
\def\ENVV{\ENV{V}} 
\def\ENVR{\ENV{R}} 
\def\ENVW{\ENV{W}} 
\def\ENVS{\ENV{S}} 

\def\ENVWC{\ENV{WC}}
\def\ENVRV{\ENV{RV}}
\def\ENVRW{\ENV{RW}}
\def\ENVRWV{\ENV{RWV}} 

\def\THIS{\code{this}}
\def\SENDER{\code{sender}}

\newcommand{\EXTEND}[2]{\ensuremath{[#1 \mapsto #2]}}

\def\partial{\ensuremath{\rightharpoonup}}

\def\PI{\ensuremath{\pi}}

\def\TRUE{\ensuremath{\textsf{T}}}
\def\FALSE{\ensuremath{\textsf{F}}}

\def\EMPTYSET{\ensuremath{\emptyset}}
\def\UNION{\cup}
\def\INTERSECT{\cap}

\def\VEC#1{\widetilde{#1}}

\def\PAR{\ensuremath{~\syntaxfont{|}~}}

\newcommand{\LSEND}[2]{\ensuremath{#1{\syntaxfont{!}}#2}}
\newcommand{\LRECV}[2]{\ensuremath{#1{\syntaxfont{?}}#2}}

\def\WHERE{\;\mid\;}

\makeatletter
\begingroup
  \catcode`|=\active
  \gdef\SET{%
    \begingroup%
    \catcode`|=\active \def|{\WHERE}
    \@ifnextchar[{\SET@args}{\SET@args[\relax]}
  }

  \gdef\SET@args[#1]#2{%
    \ifx#1\relax\relax%
      \ensuremath{\left\{\,#2\,\right\}}
    \else
      \expandafter\def\expandafter\paren@size{\csname #1\endcsname}
      \ensuremath{\paren@size\{\,#2\,\paren@size\}}
    \fi
    \endgroup%
  }

\endgroup
\makeatother

\def\LANGNAME{\textsc{TinyChain}}
\def\TINYSOL{\textsc{TinySol}}

\def\HIGH{\textsf{H}}
\def\LOW{\textsf{L}}
\def\ordleq{\ensuremath{\sqsubseteq}}

\def\TYPES{\ensuremath{\mathcal{T}}}
\newcommand{\TVAR}[1]{\ensuremath{\text{\normalfont\sffamily var}(#1)}}
\newcommand{\TCMD}[1]{\ensuremath{\text{\normalfont\sffamily cmd}(#1)}}
\newcommand{\TPROC}[3]{\ensuremath{\text{\normalfont\sffamily proc}(#1)\code{:}#2}\to#3}

\begin{document}
\maketitle
\begin{abstract}
This paper develops a model of a smart-contract language for a blockchain architecture with off-chain components.
Off-chain components are pieces of smart contracts that execute at designated locations outside of the network of blockchain nodes, but remain synchronised with the on-chain contract state.
They react to changes to the on-chain state, but may also notify the on-chain component about events in the world, e.g.\@ stock prices, weather data etc., or even act as a bridge between different blockchains.
This affords greater flexibility for the developer, but may also enable new vulnerabilities.
As a concrete example, we use the model to study the problem of ensuring integrity and secrecy of data between the on-chain and off-chain components, using static information flow control techniques.
This fails, even in the absence of a loop construct, because off-chain components act as separate threads and can encode a blocking construct e.g.\@ through recursive method calls.
We end the paper with a discussion of possible ways to remedy this situation.
\end{abstract}

\section{Introduction}
Smart contracts are programs that run atop a blockchain and are executed by all participating nodes as part of the state-update.
Originally, they were used to facilitate automatic transfer of cryptocurrency, once certain, programmable conditions were satisfied \cite{ethereum_whitepaper,ethereum_yellowpaper}.
However, with newer blockchain architectures moving away from reliance on a cryptocurrency for ensuring consensus between nodes (e.g. \cite{hyperledger_fabric_researchpaper}), smart contracts are become a general technology for coordinating distributed programs.
These blockchains are essentially distributed, replicated databases, providing a permanent record of all changes to the current state.
Such permanent, verifiable records are, for example, increasingly relevant to Additive Manufacturing (AM), also known as 3D printing, where blockchains can secure the ``digital thread'', i.e.\@ the end-to-end chain of data linking a part to its design file, material batch, machine, and process parameters, thereby underpinning certification, intellectual-property protection, and counterfeit prevention \cite{MANDOLLA2019134,kumar2025}.

Blockchain is attractive in this setting because it provides an immutable audit trail and a decentralised trust layer across distributed manufacturing stakeholders. 
However, smart contracts executing entirely on-chain are not well suited to all industrial scenarios. 
On-chain execution is ill-suited for computationally expensive tasks and for continuous interaction with events in the physical world. 
Modern blockchain architectures therefore increasingly rely on off-chain components \cite{off_chain_components}, i.e., pieces of smart-contract logic that execute at designated locations outside of the network of blockchain nodes whilst remaining synchronised with the on-chain state.
They can react to changes to the on-chain state, but may also notify the on-chain component about events in the world, e.g.\@ stock prices, weather data etc., or even act as a bridge between different blockchains.
Data storage and complex computations may thus be moved off-chain.

Off-chain components can react to blockchain updates, interface with external data sources, and support interactions with real-world systems, thereby making blockchain technology more practical for AM workflows.
This hybrid architecture, however, also raises new verification challenges.
While the blockchain offers immutability and public verifiability, off-chain components process data and perform computations beyond the direct control of the consensus mechanism.
In AM settings, such components may handle sensitive design files, production parameters, or event data originating from machines and external services, which must therefore be protected.
This is further underlined in the newly adopted EU Cyber Resilience Act (CRA) \cite{CRA}, which imposes a collection of obligations on manufacturers of products with digital elements.
Such products are specifically required to protect the \emph{integrity} and \emph{secrecy} of data, at rest or in transit \cite[Annex I, e-f]{CRA}.
Consequently, ensuring both integrity and secrecy of information flows between on-chain and off-chain components becomes essential for integrating blockchain technology with AM workflows.

Unfortunately, verifying the safety of a smart contract w.r.t.\@ any notion of safety is often hampered by the lack of a formal semantics for the smart-contract language.
Sometimes, such a semantics may be given post-hoc for the purpose of developing verification tools (see e.g.\@ \cite{grishchenko2018,schneidewind2020good_bad_ugly,eThor} for the Ethereum Virtual Machine language), but this comes with the caveat that the true specification is given by the implementation.

A different approach, followed by e.g.\@ \cite{crafa2019featherweight_solidity,bartoletti2019tinysol,chinese2021flowtypes} is to begin by defining a small calculus capturing the core features of the smart-contract language or blockchain architecture of interest, in order to study the safety properties in a more abstract setting.
In the present paper, we shall follow this latter approach.
Thus, we develop a formal model of a smart-contract language, \LANGNAME, with inline off-chain components for a blockchain architecture. 
The model is inspired by our industrial partner \cite{pbc}, but may also be used to represent other architectures with similar features; we review this architecture in Section~\ref{sec:architecture}.

The language follows in the tradition of \TINYSOL{} \cite{bartoletti2019tinysol,AGL/2024/ecoop/flowtypes,AGLM/2024/reocas/outofgas,AGL/2026/toplas/flowtypes,lybech_gorla_aceto/2026/ecoop}, which has been used to study features of the Solidity smart contract language for the Ethereum platform \cite{ethereum_yellowpaper}; however, unlike these aforementioned models, our purpose here is not to model one particular smart-contract language, but rather to model the general architecture.
It captures concurrency, point-to-point communication, and broadcast-style synchronisation between on-chain and off-chain execution contexts.
Thus, the semantics takes inspiration both from classic models of concurrent, communicating systems, such as the \PI-calculus by Milner, Walker and Parrow \cite{milner_walker_parrow1992picalc}, but also of the calculus of broadcasting systems by Prasad \cite{prasad1995broadcast}.
We give the syntax and semantics in Section~\ref{sec:syntax_semantics}.

As a concrete application, we use the model to study the problem of enforcing integrity and secrecy of information flows between on-chain and off-chain components using static information-flow control techniques, such as the classic type system by Volpano, Irvine and Smith \cite{volpano1996secure_flow_typesystem}.
In Section~\ref{sec:ifc}, we show that this fails even in the absence of an explicit loop construct, because off-chain components effectively behave as separate threads and can encode blocking behavior through recursive method calls.

\section{Architecture}\label{sec:architecture}
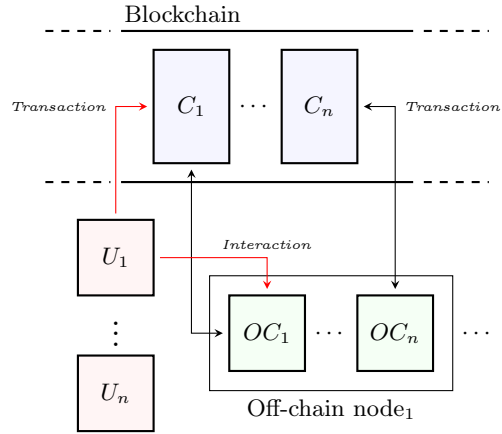
\begin{figure}\centering
\begin{tikzpicture}[
  contract/.style = { draw = black, thick, fill = blue!20!white!20,  shape = rectangle, minimum height = 1.5cm, minimum width = 1cm },
  occ/.style      = { draw = black, thick, fill = green!20!white!20, shape = rectangle, minimum height = 1cm,   minimum width = 1cm },
  user/.style     = { draw = black, thick, fill = red!20!white!20,   shape = rectangle, minimum height = 1cm,   minimum width = 1cm }
]
  \node[anchor = south west] (bc) at (0,0) { Blockchain };
  \draw[dashed, thick] (-1, 0) -- (0,0);  \draw[thick] (0,0)  -- (4,0);  \draw[dashed, thick] (4,0)  -- (5,0);
  \draw[dashed, thick] (-1,-2) -- (0,-2); \draw[thick] (0,-2) -- (4,-2); \draw[dashed, thick] (4,-2) -- (5,-2);

  \node[contract] (c1) at (1,-1) { $C_1$ };
  \node[anchor = west] (bcdots) at (c1.east) { $\Large\cdots$ };
  \node[contract, anchor = west] (cn) at (bcdots.east) { $C_n$ };

  \node[occ] (oc1) at (2,-4) { $OC_1$ };
  \node[anchor = west] (ocdots) at (oc1.east) { $\Large\cdots$ };
  \node[occ, anchor = west] (ocn) at (ocdots.east) { $OC_n$ };
  
  \node[draw = black, fit = {(oc1) (ocn)}, shape = rectangle, inner sep = 7pt, label = below:Off-chain node$_1$] (ee1) {}; 
  \node[anchor = west] (eedots) at (ee1.east) { $\Large\cdots$ };

  \node[user] (u1) at ([xshift = -2cm, yshift = 1cm]oc1) { $U_1$ };
  \node[anchor = north] (udots) at (u1.south) { $\Large\vdots$ };
  \node[user, anchor = north] (un) at (udots.south) { $U_n$ };

  \draw[<->] (oc1.west)  -| (c1.south);
  \draw[<->] (ocn.north) |- node[black, anchor = west] {\tiny\emph{Transaction}} (cn.east);

  \draw[->, red] (u1.north) |- node[black, anchor = east]  {\tiny\emph{Transaction}} (c1.west);
  \draw[->, red] (u1.east) -|  node[black, anchor = south] {\tiny\emph{Interaction}} (oc1.north);
\end{tikzpicture}
\caption{A blockchain architecture with off-chain components.}
\label{fig:off_chain_architecture}
\end{figure}

Figure~\ref{fig:off_chain_architecture} provides a high-level illustration of a blockchain architecture with off-chain components, inspired by the architecture of \cite{pbc}. 

The on-chain components (i.e.\@ smart contracts) $C_i$ reside on the blockchain itself, where they receive transactions.
Many smart-contract languages, such as Solidity \cite{ethereum_yellowpaper}, represent contracts as classes or objects in an object oriented language.
The class exposes a collection of methods, and transactions are then akin to method calls, with one extra parameter representing the \SENDER{} of the transaction, which can either be a user $U$ or another smart contract $C$.
In an architecture with off-chain components, a third possibility is that an instance of the off-chain component $OC_i$ pertaining to a contract $C_i$ may issue a transaction to $C_i$ to notify the on-chain component of the result of some computation or an event in the outside world.
A user $U$ may then either interact directly with an on-chain smart contract $C_i$ via a transaction, or alternatively indirectly via an interaction with its off-chain component $OC_i$, which may execute at one or more off-chain nodes.
In either case, the interaction pattern here is point-to-point communication between two nodes (or a user and a node).

Smart contracts are not permitted to call each other directly in the architecture of \cite{pbc}.
Instead, if contract $X$ wishes to call some method $f$ on contract $Y$, it must end the current transaction and schedule a new transaction to $Y.f$.
In order for $X$ to receive the result of that computation, $X$ may then register one or more of its own methods $f_1, \ldots, f_n$ as \emph{callbacks} for the transaction, such that when it ends, a new transaction will be generated to each of the $X.f_i$ registered callbacks.

Transactions are processed in the order in which they are received, so the blockchain executes transactions sequentially.
Whenever a transaction completes successfully, the (possibly updated) on-chain state is published to all off-chain nodes.
Each node may run off-chain components from multiple contracts, and the same off-chain component may run on multiple off-chain nodes.
Hence, the updated state is broadcast to all nodes, and the interaction pattern here is thus one-to-many communication.
Following \cite{pbc}, we view the off-chain component of a smart contract as a special method, which receives the state as a parameter and executes on a new thread at each location where it is allowed to run.

\section{The language \LANGNAME}\label{sec:syntax_semantics}
We give the syntax of \LANGNAME{} in Figure~\ref{fig:pbc_language_syntax}.
We assume the following countably infinite sets of names: $X \in \ANAMES$ (addresses); $x \in \VNAMES$ (variables); $f \in \FNAMES$ (methods); $p \in \PNAMES$ (fields); $d \in \DNAMES$ (locations, i.e.\@ blockchain or off-chain node).
We write $\VEC{\cdot}$ for a sequence of elements.


\begin{figure}
\begin{syntax}[h]
  e \in \EXPRESSIONS 
     \IS v 
     \OR x 
     \OR \THIS.p 
     \OR \op(\VEC{e}) \\
  S \in \STATEMENTS 
     \IS \SKIP 
     \OR \code{$S_1$; $S_2$} 
     \OR \DECVAR{x}{e}{S} 
     \OR \ASSIGN{e_1}{e_2}
   \ISOR \IFTHEN{e}{S_\TRUE}{S_\FALSE}
     \OR \FORK{S_1}{S_2}
   \ISOR \RPCCALL{e}{f}{\VEC{e}}{\VEC{R}}{d}
     \OR \LOCCALL{f}{\VEC{e}}{\VEC{x}} \\
  R \in \EVENTS 
     \IS \epsilon
     \OR \CALL{X}{f}{\VEC{x}}[\VEC{R}] \\
  Q \in \STACKS 
     \IS \bot 
     \OR S; Q
     \OR \ENDSCOPE{x}; Q
     \OR \ENDCALL{\ENVV, \VEC{x}, \VEC{z}}; Q 
     \OR \PUB; Q  
   \ISOR \CALLBACK{R, \VEC{x}}; Q 
     \OR \TRANS; Q \\
  T \in \TRANSACTIONS 
     \IS \TRANSACT{X_1}{X_2}{f}{\VEC{v}}{\VEC{R}} \\
  P \in \THREADPOOLS
     \IS P_1 \PAR P_2 
     \OR Q_{\ENVRV} \\
  N \in \LOCATIONS
     \IS \LOC{d}{P}{\VEC{T}}{\ENVWC}
     \OR N_1 \PAR N_2
\end{syntax}
\caption{The syntax of \LANGNAME.}
\label{fig:pbc_language_syntax}
\end{figure}


The core part of the language consists of expressions $e$ and statements $S$.
Expressions $e$ have no side effects; they are used for constant values $v$; for reading values from a locally declared variable $x$; or from the contract's state, which we model as a collection of fields $p$ that are accessed with the special keyword \THIS.  
All operations \emph{on} values (e.g.\@ addition, subtraction, multiplications etc.) are subsumed under a single construct $\op(\VEC{e})$.

Statements $S$ are the usual, imperative constructs (\code{skip}, sequential composition, local variable declaration, assignment to local variables or to the contract state, and \code{if-then}).
We have also added a forking construct $\FORK{S_1}{S_2}$ to execute $S_1$ on a separate thread whilst continuing as $S_2$.
Following \cite{pbc}, we do not include an unbounded loop construct such as \code{while}, since smart contracts are not intended to run infinite loops.
However, the language is still Turing-complete, since recursive method calls are permitted.\footnote{Many newer blockchain architectures such as Ethereum \cite{ethereum_whitepaper} feature a Turing-complete smart-contract language, but uses a limited resource, \emph{gas}, to bound the number of computational steps, cf.\@ \cite{AGLM/2024/reocas/outofgas}. We do not model gas in the present version of the language to avoid complicating the model further.}

The central constructs are the two forms of method calls, so we give some explanation of their meaning:
\begin{itemize}
  \item $\LOCCALL{f}{\VEC{e}}{\VEC{x}}$ is used for calls to local methods (i.e.\@ within the same contract), through the special variable \THIS, which is always bound to the name of the current contract.
    Unlike remote calls, these \emph{can} return results to the caller, which are passed through the out-parameters $\VEC{x}$.

  \item $\RPCCALL{e}{f}{\VEC{e}}{\VEC{R}}{d}$ is used to issue a call to another contract, and execute it at location $d$.
    The expression $e$ must yield the address of the contract; $f$ is the name of the called method; and the expressions $\VEC{e}$ are the actual arguments.    
    Lastly, $\VEC{R}$ is the \emph{event list}, which is a (possibly empty) list of \emph{callback functions}, each of which again may declare an event list.
\end{itemize}

In order to give a small-step semantics of the language, we model the execution with \emph{stacks} $Q$, following \cite{AGLM/2024/reocas/outofgas,lybech_gorla_aceto/2026/ecoop}.
Besides the empty stack $\bot$, and statements $S; Q$, we also have a collection of special stack constructs, that are central to the model:
\begin{itemize}
  \item $\ENDSCOPE{x}$ marks the end-of-scope of a locally declared variable $x$ (declared with $\DECVAR{x}{e}{S}$).

  \item $\ENDCALL{\ENVV, \VEC{x}, \VEC{z}}$ marks the end of a local method call.
    This will restore the local variable bindings $\ENVV$, that existed prior to the method call.
    Here, $\VEC{x}$ is the list of out parameters used during the execution of the function, and $\VEC{z}$ is the list of out parameters expected by the calling function, into which the values should be returned.
    Thus, it must always hold that $|\VEC{x}| = |\VEC{z}|$.

  \item $\PUB$ publishes the current state to the other locations via broadcast.
    This is used at the end of a transaction to notify the off-chain components (subscribers) about on-chain updates to the blockchain state.

  \item $\CALLBACK{R, \VEC{x}}$ is used at the end of a remote method call to generate new transactions, that invoke the functions registered as callbacks when the current function was called.
    The list $\VEC{x}$ is a list of out parameters declared in the currently executing function, and the value of these will then be passed as arguments to the functions in the event group $\VEC{R}$.

  \item $\TRANS$ takes a new transaction from the queue of transactions and places it on the stack.
    This is used at the end of a transaction call to move on to the next execution.
    We include this as a separate construct to avoid having to define different semantic rules for function execution depending on the location (since transactions are only consumed at the `blockchain' location, and not at the off-chain locations).
\end{itemize}

Transactions $T$ are of the form $\TRANSACT{X_1}{X_2}{f}{\VEC{v}}{\VEC{R}}$.
The form deliberately resembles the syntax for method invocations, since they are similar.
The only difference is that this notation also includes the caller id $X_1$, which either is a user id or a contract address.
At runtime, this will be bound to the special variable \SENDER{}, which is accessible within the body of the called method.

Each location (i.e.\@ the blockchain or the off-chain nodes) executes a \emph{thread pool} $P$, which consists of a parallel composition of thread pools $P_1 \PAR P_2$; or stacks $Q$ annotated with two environments: a \emph{read-only} memory $\ENVR$ and a local variable environment $\ENVV$.
The read-only memory is included, because state updates from the blockchain are broadcast to the off-chain nodes, where they are available to the off-chain components but cannot be modified.

Lastly, we model the system as a network of locations $N$, which consist of the blockchain and the off-chain nodes.
Thus, we represent this as a parallel composition of networks $N_1 \PAR N_2$.
A single location is written $\LOC{d}{P}{\VEC{T}}{\ENV{WC}}$ which represents a thread pool $P$, executing at the location with id $d$.
It has an associated transaction list $\VEC{T}$ (which may be empty in the case of off-chain nodes), a writable piece of shared memory $\ENVW$, and a code environment $\ENVC$.

\subsection{Operational semantics}
We can now give a semantics of \LANGNAME.
In order to model the two kinds of communication, we shall give a labelled semantics, with labels $\lambda$ given by the following syntax, where \code{!} denotes sending a message, and \code{?} denotes receiving a message:
\begin{equation*}
  \alpha  \DCLSYM \LSEND{d}{T}
          \ORSYM \LRECV{d}{T}
          \ORSYM \tau 
          \quad
  \beta   \DCLSYM \LSEND{X}{\ENVS} 
          \ORSYM \LRECV{X}{\ENVS}
          \quad
  \lambda \DCLSYM \alpha 
          \ORSYM \beta
          \ORSYM Q, \ENVRV
\end{equation*}


Traditional point-to-point communication between locations is used when one location (or a user) emits a transaction $T$, that is to be executed at a specific location $d$.
Thus, the point-to-point message labels are of the form $\LSEND{d}{T}$ and $\LRECV{d}{T}$.
We use $\alpha$ to abstract over these messages.

Broadcast messages have the form $\LSEND{X}{\ENVS}$ and $\LRECV{X}{\ENVS}$ where $X$ is a contract name (address) and $\ENVS$ is a state.
We use $\beta$ to abstract over broadcast messages.

Finally, we need a label to indicate the creation of a new thread.
We write this as simply $Q, \ENVRV$, so the label contains the entire stack of the new thread, plus the associated environment of local variable bindings $\ENVV$, and the (shared) readable memory segment, $\ENVR$.
For actions that do not generate any of the aforementioned labels, we write $\tau$ for the silent action.

The binding model for our semantics consists of three different types of environments, defined thus:
\begin{equation*}
  \begin{array}{r @{~} c @{~} l}
    \ENVC               & \in \SETENVC & \DCLSYM \ANAMES \times \FNAMES \partial \VNAMES^* \times \VNAMES^* \times \STATEMENTS  \\
    \ENVV               & \in \SETENVV & \DCLSYM \VNAMES \UNION \SET{\THIS, \SENDER} \partial \VALUES \UNION \SET{\bot} \\
    \ENVR, \ENVW, \ENVS & \in \SETENVM & \DCLSYM \ANAMES \times \PNAMES \partial \VALUES
  \end{array}
\end{equation*}

We shall use the convention of writing environments together, when they occur together.
Thus e.g.\@ $\ENVR, \ENVV = \ENVRV$.
We write $\ENV{X}\EXTEND{I}{J}$ to denote the extension of an environment $\ENV{X}$ with the binding of $I$ to $J$. 

For convenience, when indexing into the environments $\ENVC, \ENVR, \ENVW$, we write $\ENV{X}(X.i)$ for $i \in \FNAMES \UNION \PNAMES$.
Furthermore, when indexing into one of multiple environments of the same type, such as $\ENVRW$, we may write this as
\begin{equation*}
  \ENVRW(X.p) = %
  \begin{cases}
    \ENVR(X.p) & \text{if $X.p \in \dom(\ENVR)$} \\ 
    \ENVW(X.p) & \text{if $X.p \in \dom(\ENVW)$}
  \end{cases}
\end{equation*}

\noindent and otherwise undefined, and assuming $\dom(\ENVR) \INTERSECT \dom(\ENVW) = \EMPTYSET$.

The environments contain the code, locally declared variables, and the contract- and off-chain state.
\begin{itemize}
  \item $\ENVC$ is a code environment, containing the function definitions of all contracts.
    The codomain is the set of tuples $\VNAMES^* \times \VNAMES^* \times \STATEMENTS$, where such a tuple contains a list of input parameters, a list of output parameters (either or both of which may be empty), and a method body $S$.

  \item $\ENVV$ is a variable environment containing bindings for locally declared variables, including the special variables \THIS{} and \SENDER.
    Note that we use the null value $\EMPTY$ for variables that are not bound to any concrete value.
    This is the case for out parameters, when they are declared in a function definition.

  \item $\ENVR$ and $\ENVW$ are read-only and writable memories representing contract states.
    We use $\ENVS$ to refer to either of them.
\end{itemize}

The read-only memory segment $\ENVR$ represents a state that was broadcast from a different location as part of a state update.
The fact that it is read-only is not technically important, because it is not persisted once the current transaction call ends.
It could be modifiable, but any such changes would have no effect on the future state of the system.
It could therefore also be regarded as a local \emph{field environment}, analogous to the local variable environment $\ENVV$, but subject to the restriction that new entries cannot be created at runtime.

Transitions for expressions are of the form $\ENVRWV \vdash e \trans v$: they are concluded relative to the writable segment $\ENVW$ (the current state); the readable memory segment $\ENVR$; and the local variable bindings $\ENVV$. 
The semantics is standard and similar to the semantics of expressions in \TINYSOL, cf.\@ e.g.\@ \cite{AGL/2026/toplas/flowtypes}, hence we omit it here for space reasons.
Instead, we shall focus on the semantics of statements, thread pools and locations, since they are quite involved.

\subsubsection{Semantics of stacks}
The semantics of stacks is given in terms of transitions of the form 
\begin{equation*}
  \ENVC \vdash_d Q, \ENVRWV \trans[\lambda] Q', \ENVRWV' . 
\end{equation*}

Here, transitions are concluded relative to a code environment $\ENVC$; the turnstile is parametrised with a location identifier $d$, indicating the location at which the stack is executing; and the transition relation carries a label $\lambda$, which lets us observe any message generated by the transition step.

\begin{figure}\centering
\begin{irules}
  \irule{s-skip}
    { }
    { \ENVC \vdash_d \SKIP; Q, \ENVRWV \trans[\tau] Q, \ENVRWV }
    \label{rule:stm_skip}

  \irule{s-dec}
    { \ENVRWV \vdash e \trans v }
    { \ENVC \vdash_d \DECVAR{x}{e}{S}; Q, \ENVRWV \trans[\tau] S; \ENDSCOPE{x}; Q, \ENVRWV' }
    \begin{where}
      $\ENVRWV = \ENVR, \ENVW, \ENVV\EXTEND{x}{v}$
    \end{where}
    \label{rule:stm_dec}

  \irule{s-assv}
    { \ENVRWV \vdash e \trans v' }
    { \ENVC \vdash_d \ASSIGN{x}{e}; Q, \ENVRWV\EXTEND{x}{v} \trans[\tau] Q, \ENVRWV\EXTEND{x}{v'} }
    \label{rule:stm_assv}

  \irule{s-assp}
    { \ENVRWV \vdash e \trans v' }
    { \ENVC \vdash_d \ASSIGN{\THIS.p}{e}; Q, \ENVRWV\EXTEND{X.p}{v} \trans[\tau] Q, \ENVRWV\EXTEND{X.p}{v'} }
    \begin{where}
      $\ENVV(\THIS) = X$
    \end{where}
    \label{rule:stm_assp}

  \irule{s-if}
    { \ENVRWV \vdash e \trans b }
    { \ENVC \vdash_d \IFTHEN{e}{S_\TRUE}{S_\FALSE} ; Q, \ENVRWV \trans[\tau] S_b ; Q, \ENVRWV }
    [ b \in \BOOLEANS ]
    \label{rule:stm_if}

  %

  \irule{s-fork}
    { }
    { \ENVC \vdash_d \FORK{S_1}{S_2}; Q, \ENVRWV \trans[S_1;\bot, \ENVRV] S_2; Q, \ENVRWV }
    \label{rule:stm_fork}

   \irule{s-rpc}
    { \ENVRWV \vdash e \trans X \qquad \ENVRWV \vdash \VEC{e} \trans \VEC{v} }
    { \ENVC \vdash_d \RPCCALL{e}{f}{\VEC{e}}{\VEC{R}}{d'}; Q, \ENVRWV \trans[\LSEND{d'}{\TRANSACT{Y}{X}{f}{\VEC{v}}{\VEC{R}}}] Q, \ENVRWV }
    \begin{where}
      $\ENVV(\THIS) = Y$
    \end{where}
    \label{rule:stm_rpc}

  \irule{s-call}
    { \ENVRWV \vdash \VEC{e} \trans \VEC{v}_1 }
    { 
        \ENVC \vdash_d \LOCCALL{f}{\VEC{e}}{\VEC{x}}; Q, \ENVRWV 
        \trans[\tau]   S; \ENDCALL{\ENVV, \VEC{x},\VEC{z}}; Q, \ENVRWV'
    }
    \begin{where}
      \begin{math}
        \begin{array}{r @{~} l}
          \ENVV(\THIS)  & = X                                  \\
          \ENVC(X.f)    & = (\VEC{y}, \VEC{z}, S)              \\
          \ENVV         & = \ENVV''\EXTEND{\VEC{x}}{\VEC{v}_2} \\
          \ENVV'        & = \THIS:X, \SENDER:\ENVV(\SENDER), \VEC{y}:\VEC{v}_1, \VEC{z}:\VEC{v}_2
        \end{array}
      \end{math}
    \end{where}
    \label{rule:stm_call}

\end{irules}
\caption{Semantics of statements $S$.}
\label{fig:semantics_statements}
\end{figure}

First, we give semantics of the basic imperative statements $S$, which is in Figure~\ref{fig:semantics_statements}.
Most of the rules are standard for a stack-based semantics of an imperative language and resemble, to some degree, the small-step semantics of \TINYSOL{} in \cite{AGLM/2024/reocas/outofgas,lybech_gorla_aceto/2026/ecoop}.
One notable point is that this language is \emph{class-based}, with class names (addresses) $X$, and in method bodies, the special variables \THIS{} and \SENDER{} are bound (in the local variable environment $\ENVV$) to the value of the current contract, resp.\@ to the address of the caller.
The former is used when assigning to fields of the contract state, cf.\@ rule \nameref{rule:stm_assp}.
Otherwise, this rule is similar to \nameref{rule:stm_assv} for variable assignment.

In the rule \nameref{rule:stm_fork}, we simply emit a forking label $S_1; \bot, \ENVRV$ and then continue with the execution of the other statement $S_2$.
This label will be caught at the level of thread pools, where the thread will then be instantiated (cf.\@ rule \nameref{rule:thr_fork} below).
It is worth noting that the fork \emph{copies} the local variable environment $\ENVV$.
Thus, the new thread cannot communicate directly with the parent thread through the variable environment.

The main specialities are in the rules for method calls:
\begin{itemize}
  \item \nameref{rule:stm_rpc} generates a new transaction call for the current location $d$.
    This is expressed by the label $\LSEND{d}{\TRANSACT{Y}{X}{f}{\VEC{v}}{\VEC{f}}}$ which marks the \SENDER{} as the current contract, $Y$; the recipient as $X$; the method to be called is $f$ and with actual arguments $\VEC{v}$; and the set of callback methods (in the current contract) are $\VEC{f}$.

  \item \nameref{rule:stm_call} is for local method calls (in the current contract).
    This is expressed by fetching the method body $S$ from the code environment $\ENVC$, and then building the new variable environment $\ENVV'$, containing the same bindings for the special variables \THIS, \SENDER, and then also bindings for the formal parameters $\VEC{y}$, and the out-parameters $\VEC{z}$.
\end{itemize}

\begin{figure}\centering
\begin{irules}
  \irule{s-end}
    { }
    { \ENVC \vdash_d \ENDSCOPE{x}; Q, \ENVRWV\EXTEND{x}{v} \trans[\tau] Q, \ENVRWV }
    \label{rule:stm_end}

  \irule{s-ret}
    { }
    { \ENVC \vdash_d \ENDCALL{\ENVV'', \VEC{x}, \VEC{z}}; Q, \ENVRWV \trans[\tau] Q, \ENVRWV' }
    \begin{where}
      \begin{math}
        \begin{array}{r @{~} l}
          \ENVV   & = \ENVV^1\EXTEND{\VEC{z}}{\VEC{v}'} \\
          \ENVV'' & = \ENVV^2\EXTEND{\VEC{x}}{\VEC{v}}  \\
          \ENVV'  & = \ENVV^2\EXTEND{\VEC{x}}{\VEC{v}'}
        \end{array}
      \end{math}
    \end{where}
    \label{rule:stm_ret}

  \irule{s-pub}
    { }
    { \ENVC \vdash_d \PUB; Q, \ENVRWV \trans[\LSEND{X}{\ENVW}] Q, \ENVRWV }
    \begin{where}
      $\ENVV(\THIS) = X$
    \end{where}
    \label{rule:stm_pub}

  \irule{s-cb}
    {  }
    { \ENVC \vdash_d \CALLBACK{R, \VEC{z}}; Q, \ENVRWV \trans[\LSEND{d}{T}] Q, \ENVRWV }
    \begin{where}
      \begin{math}
        \begin{array}{r @{~} l}
          R         & = \TRANSACT{Y}{X}{f}{\VEC{x}}{\VEC{R}'} \\
          T         & = \TRANSACT{Y}{X}{f}{\VEC{v}}{\VEC{R}'} \\
          \ENVV     & = \ENVV'\EXTEND{\VEC{z}}{\VEC{v}}       \\
          |\VEC{z}| & = |\VEC{x}|
        \end{array}
      \end{math}
    \end{where}
    \label{rule:stm_cb}

  \irule{s-trans}
    { }
    { \ENVC \vdash_d \TRANS; Q_1, \ENVRWV \trans[\LRECV{d}{T}] Q_2, \ENVRWV' }
    \begin{where}
      \begin{math}
        \begin{array}{r @{~} l}
          T          & = \TRANSACT{Y}{X}{f}{\VEC{v}}{R_1, \ldots, R_n} \\
          \ENVC(X.f) & = (\VEC{y}, \VEC{z}, S)                \\
          \ENVV'     & = \THIS:X, \SENDER:Y, \VEC{y}:\VEC{v}, \VEC{z}:\VEC{\EMPTY} \\
          Q_2        & = S; \CALLBACK{R_1, \VEC{z}}; \ldots; \CALLBACK{R_n, \VEC{z}}; \PUB; \TRANS; \bot 
        \end{array}
      \end{math}
    \end{where}
    \label{rule:stm_trans}
\end{irules}
\caption{Semantics of stacks.}
\label{fig:semantics_stacks}
\end{figure}

The semantics of the remaining stack commands are given in Figure~\ref{fig:semantics_stacks}.
These are, in a sense, `book keeping' rules used to control the execution of stacks, such as returning from a local method call, or pulling transactions from the transaction list.

\begin{figure}\centering
\begin{irules}
  \irule{t-par$_1$}
    { \ENVC \vdash_d P_1, \ENVW \trans[\lambda] P_1', \ENVW' }
    { \ENVC \vdash_d P_1 \PAR P_2, \ENVW \trans[\lambda] P_1' \PAR P_2, \ENVW' }

  \irule{t-par$_2$}
    { \ENVC \vdash_d P_2, \ENVW \trans[\lambda] P_2', \ENVW' }
    { \ENVC \vdash_d P_1 \PAR P_2, \ENVW \trans[\lambda] P_1 \PAR P_2', \ENVW' }

  \irule{t-sub}
    { }
    { \ENVC \vdash_d P, \ENVW \trans[\LRECV{X}{\ENVS}] Q_{\ENV{SV}} \PAR P, \ENVW }
    \begin{where}
      \begin{math}
        \begin{array}{r @{~} l}
          \ENVC(X.\code{sub}) & = (\epsilon, \epsilon, S) \\ 
          \ENVV               & = \THIS:X, \SENDER:\EMPTY 
        \end{array}
      \end{math}
    \end{where}
    \label{rule:thr_sub}

  \irule{t-fork}
    { \ENVC \vdash_d P, \ENVW \trans[Q, \ENVRV] P', \ENVW }
    { \ENVC \vdash_d P, \ENVW \trans[\tau] Q_{\ENVRV} \PAR P', \ENVW }
    \label{rule:thr_fork}
\end{irules}
\caption{Semantics of thread pools.}
\label{fig:semantics_threadpools}
\end{figure}

\subsubsection{Semantics of thread pools and locations}
The semantics of thread pools $P$ are given in Figure~\ref{fig:semantics_threadpools}.
These are simply the rules for executing one of the two components in a parallel composition, and then two further rules:

\nameref{rule:thr_sub} is used for consuming a broadcast message.
Note that in this rule, we assume that every component subscribing to the broadcast messages declares a method named \code{sub}, which can receive these messages messages.\footnote{The name should be regarded as a reserved keyword. We could allow contracts to declare an arbitrary method as off-chain component, as in the architecture of \cite{pbc}, but we forego this here to avoid complicating the presentation further.}
This must be a parameterless function, declared with the body $S$, and this is then executed on a separate thread, and with the received state environment $\ENVS$ set as the read-only memory.

\nameref{rule:thr_fork} is the counterpart to the stack rule \nameref{rule:stm_fork}.
Thus, this rule is used when the executing thread is generating a new thread (at the stack level), which is then caught and inserted here at the thread pool level.
We therefore annotate the transition with $\tau$, since this is an internal action.
Therefore, the creation of local threads cannot be observed.

It is worth noting that \emph{none} of the rules describe interactions directly between threads on this level.
Each thread executes independently, but they may communicate indirectly through the shared, writable memory $\ENVW$.


\begin{figure}\centering
\begin{irules}
  \irule{l-trO$_1$}
    { \ENVC \vdash_d P, \ENVW \trans[\LSEND{d}{T}] P', \ENVW }
    { \LOC{d}{P}{\VEC{T}}{\ENVWC} \trans[\tau] \LOC{d}{P}{\VEC{T}:T}{\ENVWC} }
    \label{rule:loc_trout1}

  \irule{l-trO$_2$}
    { \ENVC \vdash_d P, \ENVW \trans[\LSEND{d'}{T}] P', \ENVW }
    { \LOC{d}{P}{\VEC{T}}{\ENVWC} \trans[\LSEND{d'}{T}] \LOC{d}{P}{\VEC{T}}{\ENVWC} }
    \begin{where}
      $d \neq d'$
    \end{where}
    \label{rule:loc_trout2}

  \irule{l-trI$_1$}
    { \ENVC \vdash_d P, \ENVW \trans[\LRECV{d}{T}] P', \ENVW' }
    { \LOC{d}{P}{T:\VEC{T}}{\ENVWC} \trans[\tau] \LOC{d}{P}{\VEC{T}}{\ENVWC} }
    \label{rule:loc_trin1}

  \nextloc{6.5,0}

  \irule{l-trI$_2$}
    { }
    { \LOC{d}{P}{\VEC{T}}{\ENVWC} \trans[\LRECV{d}{T}] \LOC{d}{P}{\VEC{T}:T}{\ENVWC} }
    \label{rule:loc_trin2}

  \irule{l-bcO}
    { \ENVC \vdash_d P, \ENVW \trans[\LSEND{X}{\ENVS}] P', \ENVW }
    { \LOC{d}{P}{\VEC{T}}{\ENVWC} \trans[\LSEND{X}{\ENVS}] \LOC{d}{P'}{\VEC{T}}{\ENVWC} }
    \label{rule:loc_bcout}

  \irule{l-bcI}
    { \ENVC \vdash_d P, \ENVW \trans[\LRECV{X}{\ENVS}] P', \ENVW }
    { \LOC{d}{P}{\VEC{T}}{\ENVWC} \trans[\LRECV{X}{\ENVS}] \LOC{d}{P}{\VEC{T}}{\ENVWC} }
    \label{rule:loc_bcin}

  \irule{l-tau}
    { \ENVC \vdash_d P, \ENVW \trans[\tau] P', \ENVW' }
    { \LOC{d}{P}{T:\VEC{T}}{\ENVWC} \trans[\tau] \LOC{d}{P}{\VEC{T}}{\ENVWC} }
    \label{rule:loc_tau}

\end{irules}
\caption{Semantics of locations.}
\label{fig:semantics_locations}
\end{figure}

In Figure~\ref{fig:semantics_locations}, we give the semantics of \emph{single} locations.
These are concerned with generating and consuming the point-to-point and broadcast communication labels.
As this is highly technical, we give some intuitions:

\nameref{rule:loc_trout1} generates a new transaction $T$ for the current location $d$ and appends it to the transaction list, cf.\@ rules \nameref{rule:stm_rpc} and \nameref{rule:stm_cb}.
This is an internal action, so the label in the conclusion is the silent action $\tau$
In contrast,  \nameref{rule:loc_trout2} generates a new transaction for a \emph{different} location $d'$ (using \nameref{rule:stm_rpc} in the premise).
Hence, the label \emph{is} observable in the conclusion.

\nameref{rule:loc_trin1} consumes a transaction $T$ from the head of the transaction list of the current location, whilst \nameref{rule:loc_trin2} receives a transaction $T$ from the outside (i.e.\@ either from a user or from a different location) and appends it to the transaction list.
Notably, this rule can always be used, since such a message can be received at any point.

\nameref{rule:loc_bcout} emits a new broadcast label (from \nameref{rule:stm_pub}); and \nameref{rule:loc_bcin} consumes a broadcast label (from \nameref{rule:thr_sub}).
Notably, the latter rule can always be used, which is important, since it ensures that locations cannot block their execution by being unable to receive a broadcast message.

At this point, let us notice that a single location can generate both broadcast messages $\LSEND{X}{\ENVS}$, $\LRECV{X}{\ENVS}$, and point-to-point communication messages $\LSEND{d}{T}$, $\LRECV{d}{T}$, as well as the internal action label $\tau$.
Recall that the former are denoted by $\beta$, and the latter (including $\tau$) are denoted by $\alpha$.
Finally, we need a semantics for \emph{networks} of locations $N$ that allow both types of communication.
We give the semantics in Figure~\ref{fig:semantics_networks}.

\begin{figure}\centering
\begin{irules}
  \irule{n-bcO$_1$}
    { N_1 \trans[\LSEND{X}{\ENVS}] N_1' \and N_2 \trans[\LRECV{X}{\ENVS}] N_2' }
    { N_1 \PAR N_2 \trans[\LSEND{X}{\ENVS}] N_1' \PAR N_2' }
    \label{rule:netw_bcOut1}

  \irule{n-bcO$_2$}
    { N_1 \trans[\LRECV{X}{\ENVS}] N_1' \and N_2 \trans[\LSEND{X}{\ENVS}] N_2' }
    { N_1 \PAR N_2 \trans[\LSEND{X}{\ENVS}] N_1' \PAR N_2' }
    \label{rule:netw_bcOut2}

  \irule{n-bcI}
    { N_1 \trans[\LRECV{X}{\ENVS}] N_1' \and N_2 \trans[\LRECV{X}{\ENVS}] N_2' }
    { N_1 \PAR N_2 \trans[\LRECV{X}{\ENVS}] N_1' \PAR N_2' }
    \label{rule:netw_bcIn}

  \nextloc{7,0}

  \irule{n-com$_1$}
    { N_1 \trans[\LSEND{d}{T}] N_1' \and N_2 \trans[\LRECV{d}{T}] N_2' }
    { N_1 \PAR N_2 \trans[\tau] N_1' \PAR N_2' }
    \label{rule:netw_com1}

  \irule{n-com$_2$}
    { N_1 \trans[\LRECV{d}{T}] N_1' \and N_2 \trans[\LSEND{d}{T}] N_2' }
    { N_1 \PAR N_2 \trans[\tau] N_1' \PAR N_2' }
    \label{rule:netw_com2}

  \irule{n-par$_1$}
    { N_1 \trans[\alpha] N_1' }
    { N_1 \PAR N_2 \trans[\alpha] N_1' \PAR N_2 }
    \label{rule:netw_par1}

  \irule{n-par$_2$}
    { N_2 \trans[\alpha] N_2' }
    { N_1 \PAR N_2 \trans[\alpha] N_1 \PAR N_2' }
    \label{rule:netw_par2}

\end{irules}
\caption{Semantics of networks.}
\label{fig:semantics_networks}
\end{figure}

\begin{itemize}
  \item The rules \nameref{rule:netw_bcOut1}, \nameref{rule:netw_bcOut2} and \nameref{rule:netw_bcIn} give the broadcast semantics of state updates, using the same technique as in the Calculus of Broadcasting Systems of Prasad \cite{prasad1995broadcast}.
    That is, output labels are repeated in the conclusion, rather than replaced by the silent action label $\tau$, thus ensuring that the output message is received by all locations in the network. 
  \item The rules \nameref{rule:netw_com1}, \nameref{rule:netw_com2} give the point-to-point communication semantics, similar to the classic \PI-calculus \cite{milner_walker_parrow1992picalc,parrow2001introduction} semantics.
  \item The rules \nameref{rule:netw_par1} and \nameref{rule:netw_par2} are used to conclude a transition for a single location.
\end{itemize}

Note that the communication and parallel composition rules only use the $\alpha$ labels, which ensures that broadcast and point-to-point communication cannot be mixed.
Thus, to conclude a $\beta$-labelled transition, every location in the network must make a transition, whereas an $\alpha$-labelled transition at most will involve two locations.

\section{Information flow control}\label{sec:ifc}
As mentioned in the Introduction, an important issue for the use of blockchain technology in many industrial settings is the ability to control the information flow between components in the system, and one of the means to achieve this is with a type system, such as the classic work by Volpano, Irvine and Smith \cite{volpano1996secure_flow_typesystem}, who created a type system with security types for ensuring integrity and secrecy of the data flows in an imperative language with statements and expressions.
The general idea is that data containers (i.e.\@ variables or fields in the present setting) are assigned a container type $\TVAR{s}$, with a security level $s$ from a lattice of security levels, e.g.\@ \HIGH{} (high) or \LOW{} (low), with $\LOW \ordleq \HIGH$.
If an expression $e$ is typed as level $s$, then the type system ensures that the \emph{highest} level of any container, that is \emph{read from} in the evaluation of $e$, is $s$ or lower.
Conversely, if a statement $S$ is typed as level $s$, then the type system ensures that the \emph{lowest} level of any container \emph{written to} within $S$, is $s$ or higher.
In other words, the type system prevents `read-up' and `write-down' of information, which ensures the dual properties of integrity and secrecy.

\begin{figure}[t]
\begin{irules}
  \irule{t-skip}
    { }
    { \Gamma \vdash \SKIP : \TCMD{s} }

  \irule{t-decv}
    { \Gamma \vdash e : s' \and \Gamma, x:\TVAR{s'} \vdash S : \TCMD{s} }
    { \Gamma \vdash \DECVAR{x : s'}{e}{S} : \TCMD{s} }

  \irule{t-assv}
    { \Gamma \vdash x : \TVAR{s} \and \Gamma \vdash e : s }
    { \Gamma \vdash \ASSIGN{x}{e} : \TCMD{s} }

  \irule{t-if}
    { \Gamma \vdash e : s \and \Gamma \vdash S_1 : \TCMD{s} \and \Gamma \vdash S_2 : \TCMD{s} }
    { \Gamma \vdash \IFTHEN{e}{S_\TRUE}{S_\FALSE} : \TCMD{s} }

  \irule{t-fork}
    { \Gamma \vdash S_1 : \TCMD{s} \and \Gamma \vdash S_2 : \TCMD{s} }
    { \Gamma \vdash \FORK{S_1}{S_2} : \TCMD{s} }

  \irule{t-call}
    { \Gamma \vdash \THIS.f : \TPROC{\VEC{s}_1}{\VEC{s}_2}{s} \and \Gamma \vdash \VEC{e} : \VEC{s}_1 \and \Gamma \vdash \VEC{x} : \TVAR{\VEC{s}_2} }
    { \Gamma \vdash \LOCCALL{f}{\VEC{e}}{\VEC{x}} : \TCMD{s} }

  \irule{t-rpc}
    { \Gamma \vdash e.f : \TPROC{\VEC{s}}{F}{s} \and \Gamma \vdash \VEC{e} : \VEC{s} \and \Gamma \vdash \VEC{R} : F }
    { \Gamma \vdash \RPCCALL{e}{f}{\VEC{e}}{\VEC{R}}{d} : \TCMD{s} }

  \nextloc{4.5,0}
  \irule{t-seq}
    { \Gamma \vdash S_1 : \TCMD{s} \and \Gamma \vdash S_2 : \TCMD{s} }
    { \Gamma \vdash S_1; S_2 : \TCMD{s} }

  \nextloc{6,-2.5}
  \irule{t-assp}
    { \Gamma \vdash \THIS.p : \TVAR{s} \and \Gamma \vdash e : s }
    { \Gamma \vdash \ASSIGN{\THIS.p}{e} : \TCMD{s} }

\end{irules}
\caption{Type rules for statements}
\label{fig:typerules_stm}
\end{figure}

Variants of this type system have already been developed for \TINYSOL{} \cite{AGL/2024/ecoop/flowtypes,AGL/2026/toplas/flowtypes,lybech_gorla_aceto/2026/ecoop}.
Likewise, in the architecture of \cite{pbc}, a simple variant of this type system (with just two levels) is employed to prevent data flows from secret to public variables in their smart-contract language.
However, unlike in the \TINYSOL{} setting, that language has both implicit and explicit \emph{concurrency}, just as our language \LANGNAME{} does; namely implicitly because each off-chain component is executed on a new thread whenever a state update is received; and explicitly via the $\FORK{S_1}{S_2}$ command.
This can be problematic w.r.t.\@ ensuring secrecy, because concurrency may allow information to be leaked from higher to lower level variables via a form of side-channel attack, as Smith and Volpano \cite{SV98} demonstrated for their calculus extended with threads.
Their example relied on threads communicating via shared variables, and an unbounded loop construct, \code{while}, which was used to cause threads to block.
Neither of these features are present in \LANGNAME, but even in their absence, we can show that a direct adaptation of the security type system of \cite{volpano1996secure_flow_typesystem} for single-threaded programs cannot be used to preserve secrecy in our language.

Consider the following language of types:
\begin{syntax}[h]
  \tau \in \TYPES
       \IS s \OR \TVAR{s} \OR \TCMD{s} \OR \TPROC{\VEC{s}}{F}{s} \\ 
  F    \IS \TPROC{\VEC{s}}{F}{s} \OR \VEC{s}
\end{syntax}

Types are given to values, expressions and statements as described above; and also to methods, which we give the type $\TPROC{\VEC{s}}{F}{s}$.
This should be read as: given that the actual arguments have types $\VEC{s}$, and the out-parameters resp.\@ callbacks have types $F$, then the body of the method can be typed as $\TCMD{s}$.
We use a type environment $\Gamma$, defined as 
\begin{equation*}
  \Gamma \DCLSYM \VNAMES \UNION \ANAMES \times \PNAMES \UNION \ANAMES \times \FNAMES \partial \TYPES
\end{equation*}

\noindent to record the type assumptions of contract names, variables, fields and methods.
We assume contract declarations are annotated with types, and contracts are well-typed if the bodies of all its methods are typable according to the method signatures and type constraints of declared fields.
The type rules for statements $S$ are given in Figure~\ref{fig:typerules_stm}.\footnote{We omit type rules for expressions $e$, as well as subtyping rules, since they are not needed for the presentation, but these can easily be adapted from e.g.\@ \cite{AGL/2024/ecoop/flowtypes}. Rules for stacks, thread pools, locations, transactions and the message labels would also be needed, if one were to show confinement and non-interference, but as we know these properties cannot hold, we shall omit the rules.}

Now, consider the following contract declaration:
\begin{lstlisting}[style = tinysol]
contract X {
  field x := @$v$@ : @$\TVAR{\HIGH}$@, y := @$\EMPTY$@ : @$\TVAR{\LOW}$@, z := @$\EMPTY$@ : @$\TVAR{\LOW}$@

  func sety(val : @$\LOW$@) { this.y := val }

  func sub():@$\epsilon\to\LOW$@ {
    call this.block(this.x, this.y);
    this.z = this.y
  }

  func block(h : @$\HIGH$@, l : @$\LOW$@):@$\epsilon\to\LOW$@ {
    if (h @$\neq$@ l) then call this.block(h, l) else skip
  }
}
\end{lstlisting}

We assume the fields \code{x,y,z} are used to store only boolean values.
The field \code{x} contains a High (i.e.\@ secret) value $v$, and the two other fields are uninitialised.
Now suppose we schedule two transactions from an arbitrary user $U$:
\begin{equation*}
  \TRANSACT{U}{X}{\code{sety}}{\TRUE}{\epsilon}  \qquad\text{and}\qquad
  \TRANSACT{U}{X}{\code{sety}}{\FALSE}{\epsilon}
\end{equation*}

Both will trigger an update of the contract state, which will execute the off-chain component at any off-chain node, cf.\@ rules \nameref{rule:stm_pub} and \nameref{rule:thr_sub}.
In both cases, this invokes the \code{block} method, which blocks if the two values are unequal, by recursively calling itself with the same values, and otherwise does nothing.
Only in the case where the value in \code{y}, which is set by the transaction, equals the secret value $v$, will the thread running the off-chain component be able to make progress and assign the value of \code{y} to \code{z}.
Thus, the secret value of \code{x} is leaked to the public field \code{z}, in violation of the secrecy property intended by the types.

The above example works, because the set of boolean values has exactly two members.
With e.g.\@ integers, this approach would require scheduling infinitely many transactions, which of course is not feasible.
However, if integers were permitted to be read as bit vectors, then it would still be possible to guess the secret value bit by bit, as in the example used by Smith and Volpano in \cite{SV98}.

In that work, Smith and Volpano also proposed a solution to the problem:
It consisted in requiring both the guard $e$ and body of all \code{while} loops to be typable as $\LOW$ resp.\@ \TCMD{\LOW}, which prevents the them from being used directly in a side-channel attack.
Unfortunately, that solution is not as easily applicable in the present setting, where we used a recursive method call, since the language lacks a loop construct.
The equivalent would be to require all arguments and method bodies to be typable as $\TCMD{\LOW}$, which effectively would disallow any modification of secret values at all, which clearly is too restrictive.
Alternatively, we could forbid unbounded recursive method calls, which however would limit the expressive power of the language.
Clearly, neither option seems attractive.

Another alternative could be to use a counting technique, similar to the one used in \cite{AGLM/2024/reocas/outofgas} to statically bound the use of recursive method calls and thereby prevent infinite recursions.
This could be combined with the above solution, such that unbounded recursions might still be allowed, \emph{if} it neither depends on, nor modifies, any secret values.
In our future work, we intend to explore whether this is a viable solution.

\section{Conclusions and future work}
In the present paper, we have presented the calculus \LANGNAME, which models a blockchain architecture with off-chain components.
The model features transactions, threads and locations; and communication between location which may be used to represent both blockchains and off-chain components.
Even multiple blockchains may be represented, in case one wishes to model a multi-layer blockchain system, such as is possible with e.g.\@ Hyperledger Fabric \cite{hyperledger_fabric_researchpaper}.
The semantics mixes point-to-point and broadcast communication between locations, and is therefore quite involved to work with.
Hence, to facilitate its use, we intend to build a formalisation in a proof assistant, as part of our future work.

The purpose of the model is to serve as a formal foundation for reasoning about security properties in more complex blockchain systems, which, unlike similar previous studies \cite{AGL/2024/ecoop/flowtypes,AGL/2026/toplas/flowtypes,AGLM/2024/reocas/outofgas,lybech_gorla_aceto/2026/ecoop} may consist of multiple components, executing concurrently.
In the present paper, we have focused on integrity and secrecy, because of their direct relevance in the Additive Manufacturing domain, following the introduction of the Cyber Resilience Act \cite{CRA}, which explicitly mentions these properties. 
Our example shows that these properties cannot be ensured with a direct application of the usual type system for information flow control. 
This highlights the benefit of having a formal model of the system to study its security properties, before an implementation is created.
As part of our future work, we intend to explore whether integrity and secrecy can be recovered without severely reducing the expressive power of the language.

\begin{credits}
\subsubsection{\ackname} 
This work was supported by the SECUREAM project, DIREC, NFC, Industriens Fond under Grant No. 25808.

\subsubsection{\discintname}
The authors declare that they have no competing interests.
\end{credits}

\bibliographystyle{splncs04}
\bibliography{literature}

\end{document}